\documentclass[aps,prl,twocolumn,showpacs]{revtex4}

\usepackage{graphicx}
\usepackage{amsmath,amssymb}

\begin{document}


\title{Periodic Orbits and Binary Collisions 
in the Classical Coulomb Three-Body Problem}


\author{Mitsusada~M.~Sano}
\affiliation{
Graduate School of Human and Environmental Studies, Kyoto University, 
Sakyo-ku, Kyoto, 606-8501, Japan
}

\author{Kiyotaka~Tanikawa}
\affiliation{
National Astronomical Observatory of Japan, 
Mitaka, Tokyo, 181-8588, Japan
}


\date{\today}

\begin{abstract}
In the helium case of the classical Coulomb three-body problem 
in two dimensions with zero angular momentum, 
we develop a procedure to find periodic orbits applying 
two symbolic dynamics for one-dimensional and planar problems. 
A sequence of periodic orbits are predicted and 
are actually found numerically. 
The results obtained here will be a cornerstone for finding 
the remaining periodic orbits, 
which needed for semiclassical applications 
such as periodic orbit quantization. 
\end{abstract}


\maketitle


The microscopic three-body problem, for example, the helium atom, 
attracts both theoreticians and experimentalists 
because of rich structure of its spectrum. 
In particular, the excited states 
below the double ionization threshold ($E=0$)  
show the most complicated spectral structure, and 
contain various information on electron-electron correlations. 
This spectral region is not only most interesting 
but also most difficult in both theoretical and experimental aspects. 
In investigating its spectrum, semiclassical methods inevitably face 
with chaotic nature of its classical dynamics~\cite{TRR}. 
In a theoretical aspect, for the collinear {\it eZe} configuration, 
the semiclassical periodic orbit quantization was carried out 
by using the celebrated Gutzwiller trace formula~\cite{Gutzwiller} 
within reasonable accuracy~\cite{ERTW,TW}.
However, such successes of semiclassical and classical approaches 
have been restricted to one-dimensional case, namely 
the collinear {\it eZe} and {\it eeZ} configurations  
and the Wannier ridge configuration~\cite{TRR,ERTW,TW,RTW}. 
The helium atom in two dimensions remains as a challenging object. 
 
Main difficulties are due to (1) the mixture of chaos and tori, 
(2) the high-dimensionality,  
and (3) the singular nature of its dynamics. 
To overcome these difficulties, 
one needs sophisticated mathematical tools. 
McGehee's blow-up transformation 
and the concept of the triple collision manifold~\cite{McGehee} 
are important examples~\cite{BGY,Sano1,Sano2,Sano3,CLT,LTN,LCT}. 
Thanks to these tools, 
the dynamics near triple collision and the structure of stable 
and unstable manifolds were elucidated to some extent. In fact, 
a quantum manifestation of triple collision was 
demonstrated~\cite{BCLT,TCLCD}. 
But we still know less about periodic orbits 
in two-dimensional dynamics, 
which are needed for semiclassical analysis. 

Quite recently, one additional tool, i.e., the symbolic dynamics
for the planar gravitational problem is developed~\cite{TM}. 
In this method, binary collisions are naturally 
obtained as boundaries of different symbol sequences. 
We already have one-dimensional symbolic dynamics~\cite{Sano1}. 
Then, we can systematically search periodic orbits in the dynamics 
with zero initial velocities (DZIV) using one-dimensional and 
two-dimensional symbolic dynamics. We do this in the present report. 

As is known in \cite{Sano3}, the initial condition space 
of the DZIV is bounded by the collinear initial conditions (see below). 
First, using the two-dimensional symbolic dynamics,  
we numerically confirm the existence of curves of initial conditions 
of orbits exhibiting binary collisions. We call these  
the binary-collision curves (BCCs). 
Next, we go back to symbolic dynamics for the collinear problem and 
obtain symbol sequences of triple collision orbits. 
As predicted in \cite{Sano3}, a sequence of BCCs is shown to have two 
end-points on $\alpha = \pi$. 
In other words, these BCCs can be uniquely specified 
by the end-points. Then, we informally consider a one-dimensional
symbolic dynamics along BCCs and obtain self-retracing periodic orbits 
along them.  
Our procedure (Fig.~\ref{fig3}) will be justified 
by numerical integration (Fig.~\ref{fig4}). 
Finally, we characterize periodic orbits with physical quantities. 

Now we introduce the system considered. In the helium atom, 
two electrons are denoted by particles $1$ and $2$, and the nucleus is 
denoted by particle $3$. Let $\boldsymbol{r}_{i}$ be the position vector 
of the $i$th particle, and let  
$\boldsymbol{r}_{ij}=\boldsymbol{r}_{i}-\boldsymbol{r}_{j}$. 
We assume that the nucleus has infinite mass and that initial velocities
of particles are zero, which implies that the problem is planar. 
Then, the Hamiltonian in the hyperspherical coordinates is given by 
\begin{eqnarray}
H & = & \frac{1}{2} \left ( p_{r}^{2} + \frac{4p_{\chi}^{2}}{r^{2}} 
+ \frac{4p_{\alpha}^{2}}{r^{2}\sin^{2}(\chi)} \right )
\nonumber \\
& & 
- \frac{1}{r} \left ( 
\frac{1}{\cos\left ( \frac{\chi}{2} \right )} 
+ \frac{1}{\sin\left ( \frac{\chi}{2} \right )} 
- \frac{1}{Z[1-\sin(\chi)\cos(\alpha)]^{1/2}}
\right ),
\label{eq:hamiltonian_hyper}
\end{eqnarray}
where $r=(r_{1}^{2}+r_{2}^{2})^{1/2}$ is the hyperradius with 
$r_{1} = r \cos(\chi/2)$, $r_{2}= r \sin(\chi/2)$, 
and $\chi=2 \mbox{arctan}(r_{2}/r_{1})$. 
$Z$($=2$ in this Letter) is the charge of the nucleus. 
$Z$ appears in the denominator of the potential term because of 
the scaling of variables. $\alpha$ is the angle between the vectors 
$\boldsymbol{r}_{13}$ and $\boldsymbol{r}_{23}$. 
The total energy is $E=H$. 

Next we introduce symbolic dynamics~\cite{TM}.
Three particles from a triangle. The area of this triangle 
has the sign $\eta = {\bf e}_{z}\cdot (\boldsymbol{r}_{13}\times 
\boldsymbol{r}_{23})/|\boldsymbol{r}_{13}\times\boldsymbol{r}_{23}|$. 
The length of the $i$th side of the triangle is $l_{i} = |\boldsymbol{r}_{jk}|$, 
where $(i,j,k)$ is a cyclic permutation of $(1,2,3)$. If $\eta$ changes
its sign from plus to minus and $\mbox{max}\{l_{1},l_{2},l_{3}\}=l_{k}$, 
then we assign a symbol $k$. If $\eta$ changes its sign from minus 
to plus and $\mbox{max}\{ l_{1},l_{2},l_{3}\}=l_{k}$, 
then we assign a symbol $k+3$. 
Then we obtain a symbol sequence ${\bf s}= \bullet s_{1}s_{2}s_{3}\cdots$ 
for a given orbit with the symbol set ${\bf S}=\{1,2,3,4,5,6\}$. 

\begin{figure}
\begin{center}
\includegraphics[width=8cm]{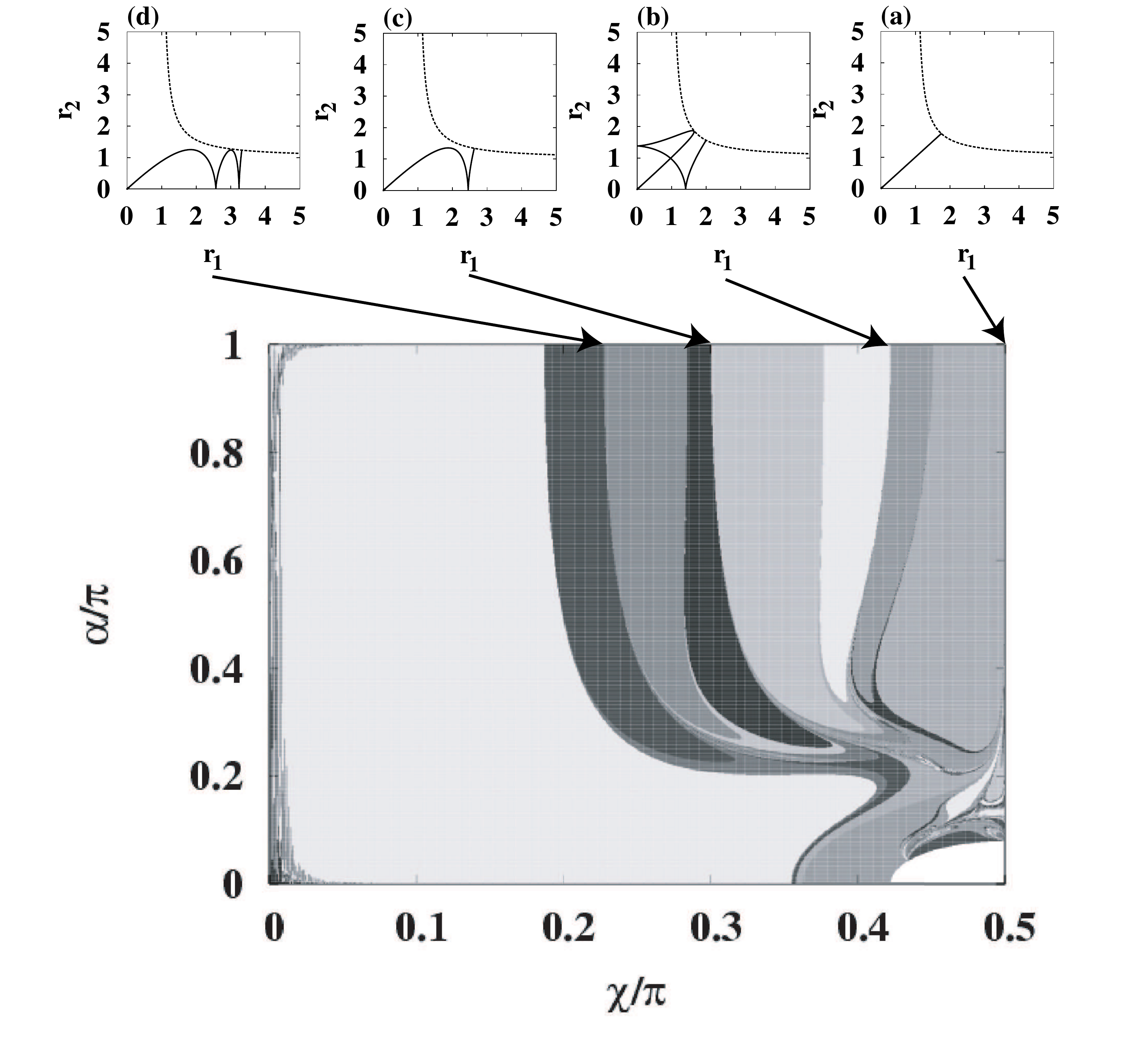}
\end{center}
\caption{\label{fig1} Partition in $D_{1/4}$
and triple-collision points in the collinear {\it eZe} configuration: 
The large figure represents the partition obtained 
from the symbol sequences of symbol length $8$ 
in the symbolic dynamics for two-dimensional dynamics. 
The boundaries between the regions of the partition are 
binary-collision curves. 
The small figures represent the triple-collision orbits 
in the collinear {\it eZe} configuration. 
(a) $0$. (b) $210$. (c) $20$. (d) $220$. 
The blank region at the lower-right corner 
is the forbidden region ($E>0$). 
}
\end{figure}

By scaling the system, we remove the dependence on $r$ from the DZIV.  
Then the initial conditions are uniquely specified by 
$0 \leq \chi < \pi$ and $0 \leq \alpha < 2\pi$. 
Moreover, the quarter of the initial condition space 
$D_{1/4}= \{ (\chi,\alpha)| 0\leq \chi \leq \frac{\pi}{2}, 
0 \leq \alpha \leq \pi \}$ suffices for our purpose owing to 
the symmetry of the problem. 
We integrate orbits starting at points in $D_{1/4}$, and assign 
a symbol sequence to each orbit following the procedure given in 
\cite{TM}. Then, $D_{1/4}$ is partitioned into regions of different 
symbol sequences. One example of partitions by symbol sequences 
up to symbol length $8$ is shown in Fig.~\ref{fig1}. 
The BCCs are clearly seen as the boundaries of the partitioned 
regions. This shows a big advance compared with \cite{Sano3}. 
In that work, BCCs were obtained point by point along them, so the task 
was rather time consuming. In addition, orbit integrations near 
$\alpha=\pi$ were difficult due to the closeness to triple collision.  
The present method basically avoids triple collision, so is accurate 
and fast. 

The intersections of BCCs with $\alpha = \pi$ are the initial conditions 
for triple-collision orbits (TCOs). 
We call them the triple-collision points (TCPs). 
We attach, in Fig.~\ref{fig1}, the orbits of some of the TCPs 
with symbol sequences $0$, $210$, $20$, and $220$ where '$0$' represents 
triple collision, '$1$' binary collision between particles $1$ and $3$, 
and '$2$' binary collision between particles $2$ and $3$. 
Hereafter the symbolic dynamics means the symbolic dynamics 
in the collinear {\it eZe} configuration. 

We observe two groups of BCCs. 
One group, which we call the {\it first kind}, 
consists of the BCCs which start at 
$\alpha = \pi$ and end at $\alpha = \pi$. 
We denote a BCC of this group by a lobe from its shape, 
and express it as $w_m0$-$w_n0$ using the symbol 
sequences $w_m0$ and $w_n0$ of the left and right end points
where $w_k$ is a word of length $k$. 
The other, which we call the {\it second kind}, 
consists of the BCCs which start at $\alpha =\pi$ and end at $\alpha = 0$. 
There are other possibilities of the terminals of the BCCs. 
For our present purpose, we only consider periodic orbits on BBCs of the
first kind.  
There exist an infinite number of TCPs on $\alpha = \pi$. 
Correspondingly, the number of lobes is infinite
(see also Fig.~\ref{fig5} and~\cite{Sano1}), which immediately 
implies that there exist infinitely many (unstable) POs on BCCs of 
the first kind, as we show later. 

\begin{figure}
\begin{center}
\includegraphics[width=7cm]{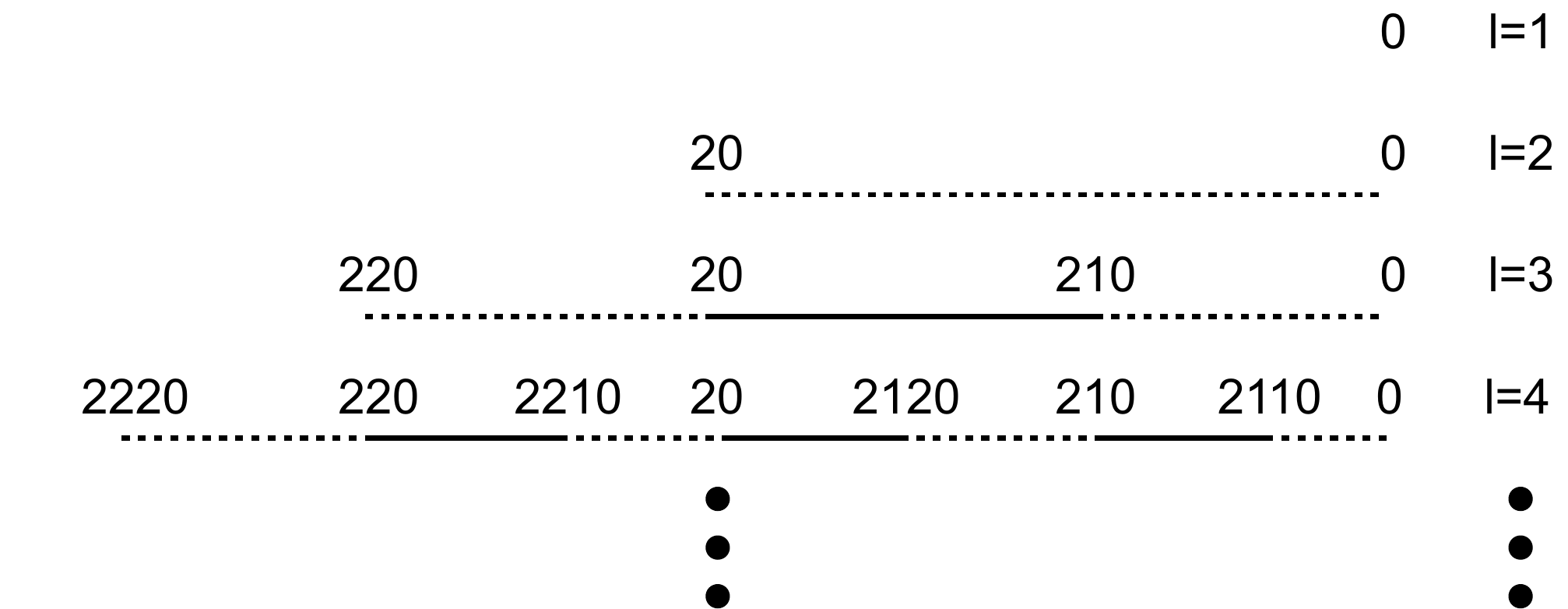}
\end{center}
\caption{\label{fig2} 
Appearance of triple-collision points on the line 
$\alpha = \pi$: $l$ is the symbol length in the symbolic dynamics 
for the collinear {\it eZe} dynamics. 
The symbol sequences of triple-collision orbits on $\alpha = \pi$ 
are represented in the symbolic dynamics 
of the collinear {\it eZe} case. 
The solid (resp. dotted) line represents the lobe corresponding 
to $2$-$3$ (resp. $1$-$3$) collision. 
}
\end{figure}

The appearance order of TCPs is illustrated in Fig.~\ref{fig2}. 
If the symbol length is $l$, the total number of the TCPs 
appearing on $\alpha = \pi$ is $2^{l-1}$ where 
$l$ denotes the {\it generation} of TCPs. 
The successive generations of TCPs are connected by the following rule:
In the $(l+1)$th generation, new TCPs are added in between 
the already existing TCPs in the $l$th generation. 
Let $w$ be a word of length $l-1$, and $w0$ be a symbol sequence of 
a TCP which already exists in the $l$th generation. 
Then, $w10$ appears to the immediate right of $w0$, 
while $w20$ appears to the immediate left of $w0$. 
(The former rule does not apply for $l=1$.) 
Any orbit in BCCs with this new $w10$ or $w20$ at the right 
(resp. left) end experiences a 2-3 (resp. 1-3) collision. 

\begin{figure}
\begin{center}
\begin{tabular}{c}
\includegraphics[width=7cm]{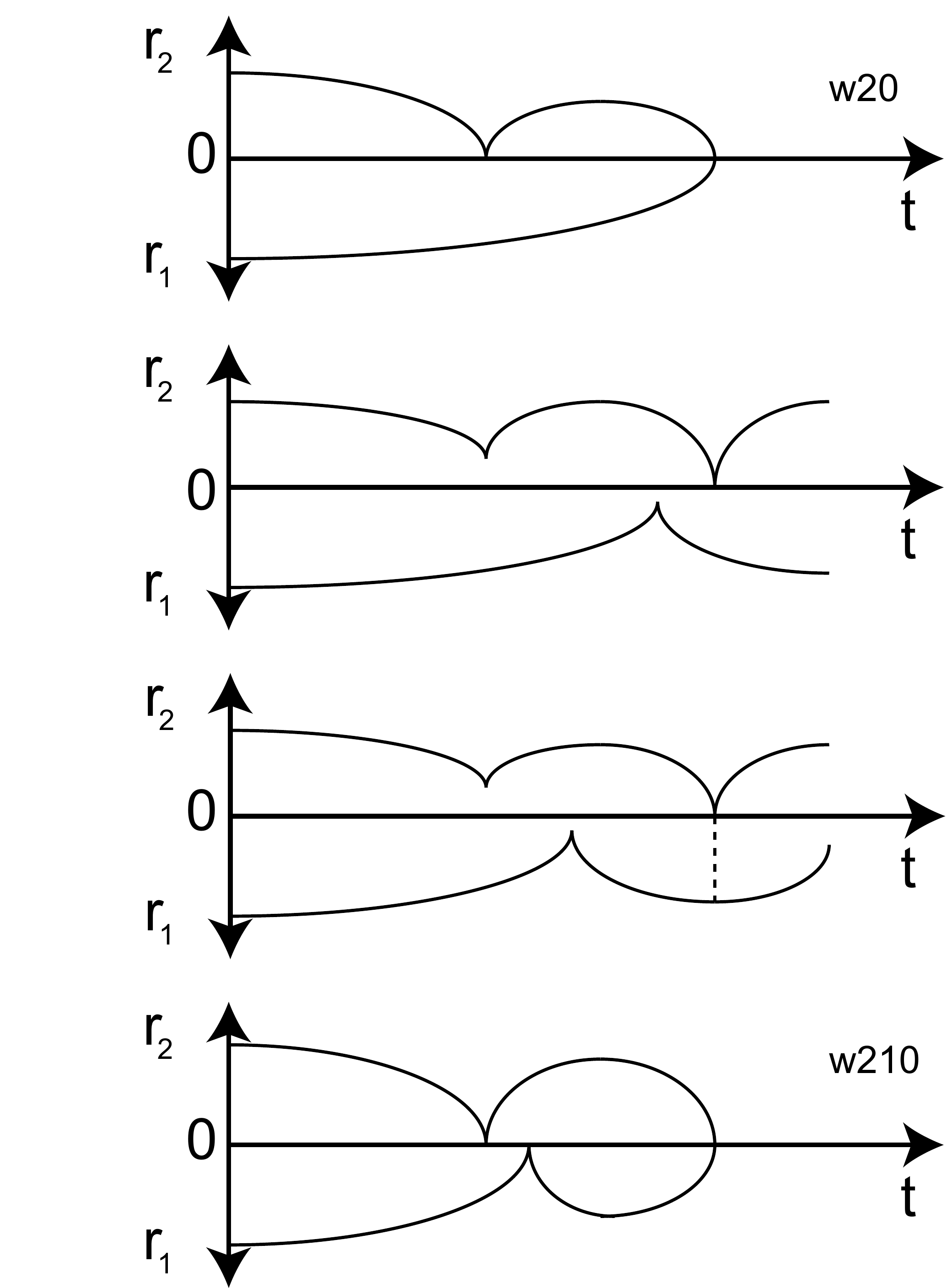}
\end{tabular}
\end{center}
\caption{\label{fig3} Tails of orbits on the lobe $w20$-$w210$
}
\end{figure}

\begin{figure}
\begin{center}
\begin{tabular}{cc}
\includegraphics[width=3.5cm]{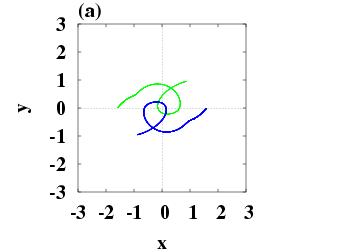} &
\includegraphics[width=3.5cm]{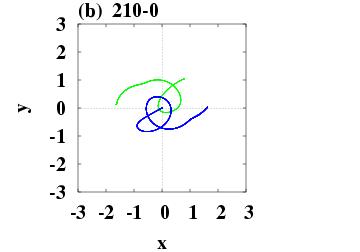} \\
\includegraphics[width=3.5cm]{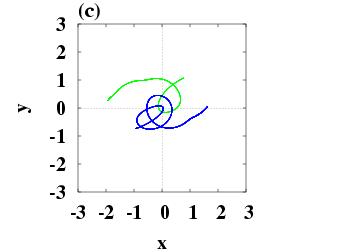} &
\includegraphics[width=3.5cm]{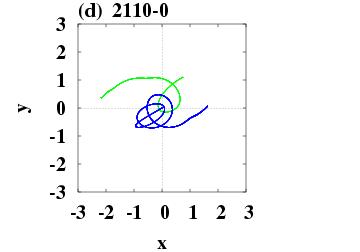} \\
\includegraphics[width=3.5cm]{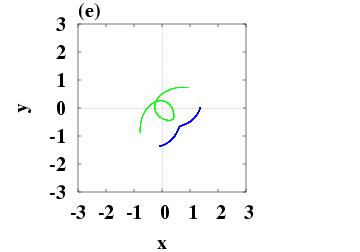} &
\includegraphics[width=3.5cm]{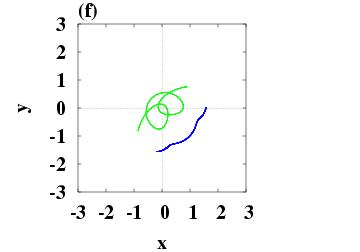} \\
\includegraphics[width=3.5cm]{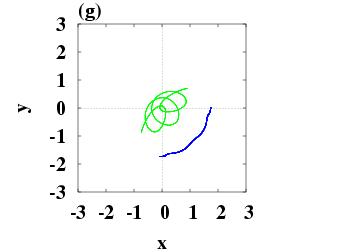} &
\includegraphics[width=3.5cm]{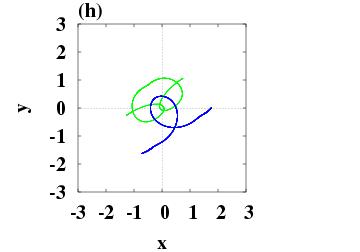} \\
\includegraphics[width=3.5cm]{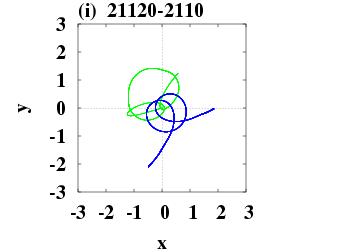} &
\includegraphics[width=3.5cm]{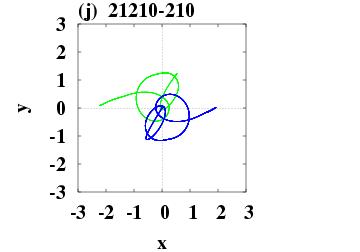} 
\end{tabular}
\end{center}
\caption{\label{fig4} (Color online) 
Periodic orbits (POs) found~: 
The nucleus is fixed at the origin. 
For the POs with binary collisions, 
names of the lobe to which they belong are indicated. 
In the orbit integration, we set $E=-1$. 
}
\end{figure}

Here we give an intuitive proof of the existence of a PO on each BCC 
of the first kind (i.e., lobe). We introduce coordinate $h$ 
($0 \leq h \leq 1$) on the curve so that $h=0$ at the left end and 
$h=1$ at the right end. The orbit changes  
its topology when we move along the curve from $h=0$ to $h=1$ 
since the orbits at both ends have different histories of collisions. 
There are three types of lobes: 
(i) lobe $w20$-$w0$ ($w \neq 2^k$, $k \geq 0$) or lobe $w0$-$w10$; 
(ii) lobe $w0$-$w'0$ where symbol lengths of $w$ and $w'$ differ by 
2 or more; 
(iii) lobe $2^{k+1}0$-$2^{k}0$ ($k\geq 0$).

First consider case (i), i.e., the case of lobe $w20$-$w210$. 
Let us move from $w20$ ($h=0$) to $w210$ ($h=1$). 
The relative position of the minimum of $r_{1}$ continuously changes 
as in Fig.~\ref{fig3}. 
The triple collision at $h=0$ becomes a $2$-$3$ binary collision 
for $ 0 < h < 1$. As $h$ increases, the next minimum of $r_1$ approaches 
the $2$-$3$ binary collision from the right, and finally coincides with 
it at $h=1$ resulting in triple collision. 
In this process, there exists a parameter value  
such that the $2$-$3$ binary collision and the local maximum of $r_{1}$
take place at the same time (see the third panel of Fig.~\ref{fig3}). 
At this moment, the angular momentum of electron 2 and the nucleus 
is zero with respect to their center of mass, which means that the angular 
momentum of electron 2 is zero, since the nucleus has infinite mass. 
This in turn means that the angular momentum of electron 1 is zero, 
which implies that electron 1 stands still at this moment. 
Then, both electrons retrace the path they tread, that is, 
a self-retracing PO is obtained. In the above proof, the persistence of the 
minima of $r_1$ and $r_2$ is crucial. 

The proof for case (ii) may be similar to case (i). 
However, there is a possibility that the orbit escapes, that is, one of 
electrons escapes to infinity. We skip this case, since we need a
special care to treat the non-persistence of minima of $r_1$ and/or $r_2$. 

In case (iii), the situation is different from that of case (i).   
In this case, we can draw a similar orbital change as in Fig.~\ref{fig3}.
This time, however, the third minimum of $r_2$ disappears due to the 
escape of the orbit when the first minimum approaches the second 
minimum of $r_2$. Then we do not have the simultaneous occurrence of 
the maximum of $r_2$ and the minimum of $r_1$. 
Therefore, there is no POs on the $2^{k+1}0$-$2^{k}0$ lobes.

\begin{figure}
\begin{center}
\begin{tabular}{c}
\includegraphics[width=7cm]{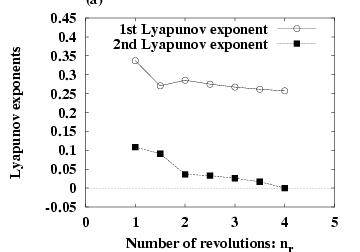}\\
\includegraphics[width=7cm]{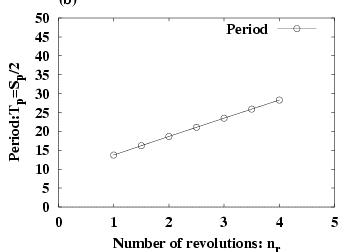} 
\end{tabular}
\end{center}
\caption{\label{fig5} 
The Lyapunov exponents and the period for a family of periodic orbits: 
(a) The Lyapunov exponents. (b) The period. 
$n_{r}$ is the number of revolutions. 
}
\end{figure}
\begin{figure}
\begin{center}
\begin{tabular}{c}
\includegraphics[width=6cm]{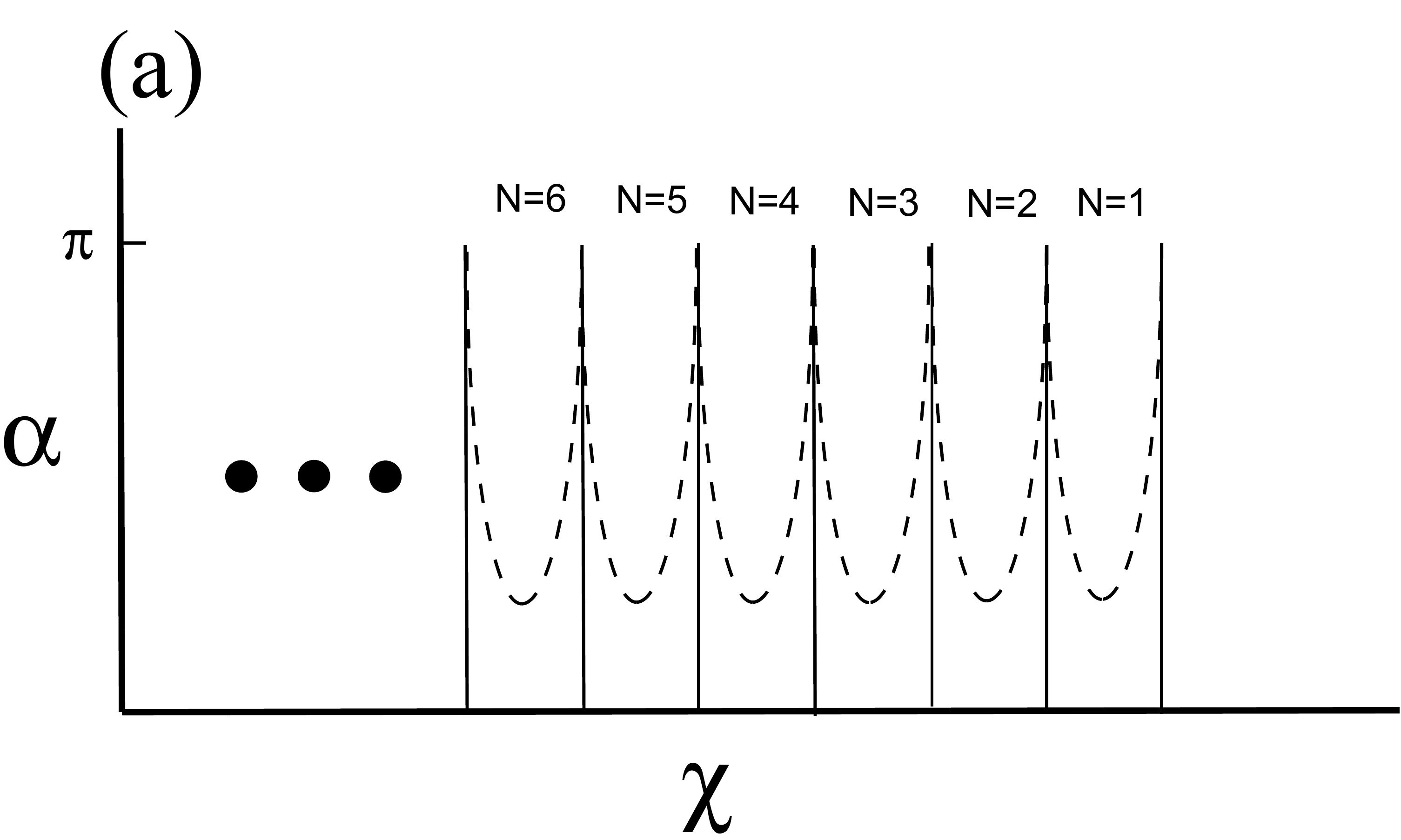} \\
\includegraphics[width=6cm]{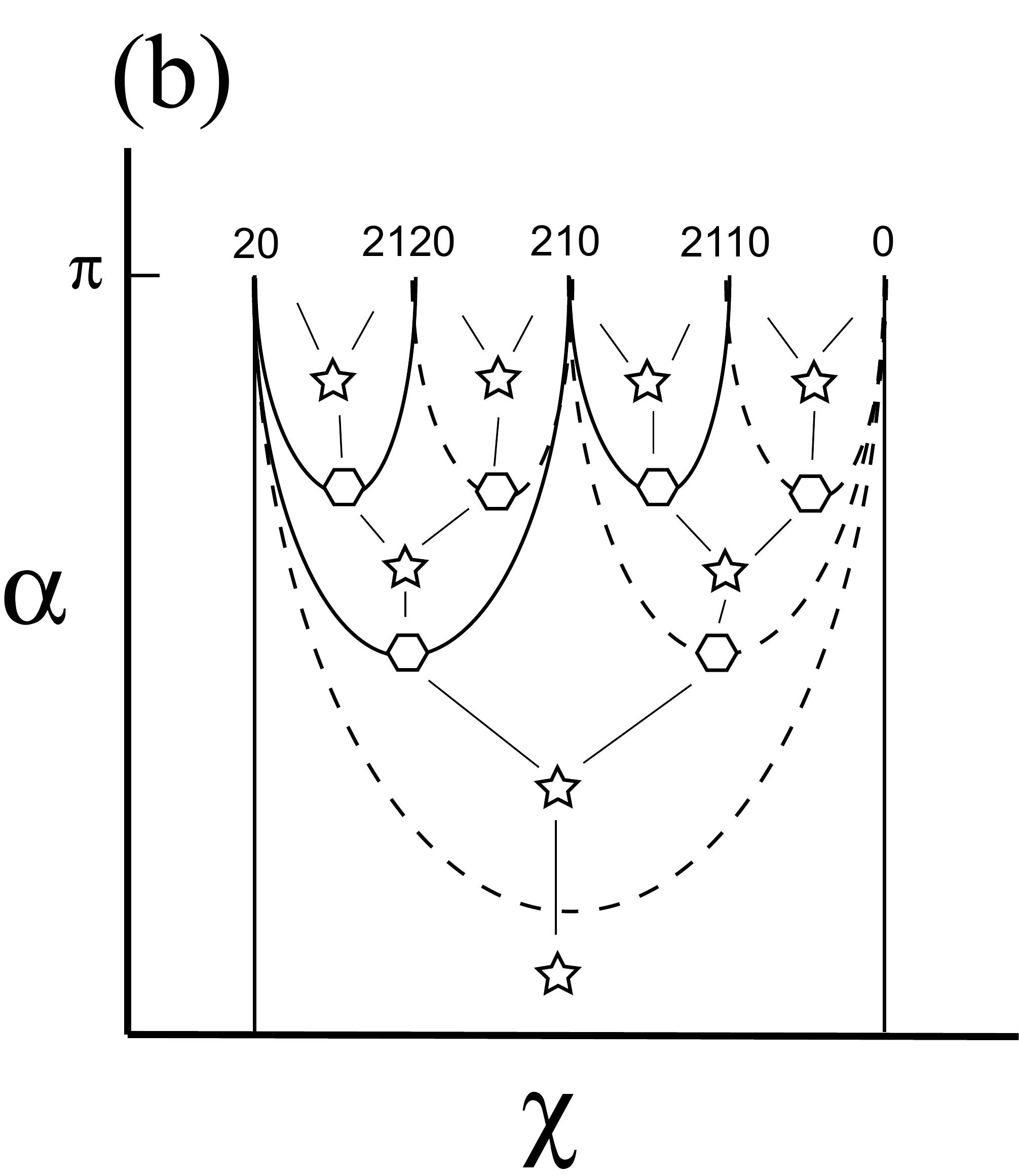} 
\end{tabular}
\end{center}
\caption{\label{fig6} Schematic pictures of binary-collision curves 
and periodic orbits (POs) in $D_{1/4}$:
(a) Schematic picture of binary-collision curves (BCCs) of the first kind and 
the second kind. Solid (resp. dotted) lines represent the BCCs 
for $2$-$3$ (resp. $1$-$3$) collisions. 
There are infinite number of fundamental blocks. 
The fundamental block is numbered by $N$ from the right hand side. 
(b) POs in the fundamental block $N=1$: 
The triple collision orbits in the line $\alpha = \pi$ are indicated. 
The BCCs of the first kind are nested. 
The hexagon (resp. star) represents a position of a PO with (resp. without) 
binary collisions. 
The connected graph, which links the positions of the POs, is displayed. 
}
\end{figure}

To obtain POs numerically, 
we look for the moment of zero velocities 
other than the initial moment. 
We actually carried out this procedure
with many trials and errors. 
Then we find some POs using the improved Newton method. 

We show example trajectories of POs in Fig.~\ref{fig4}. 
We note one remarkable feature in these POs. One electron, 
by rapidly revolving round the nucleus, screens (or weakens) 
the Coulomb field which the other electron feels. 
During this phase, the other electron moves slowly. 
Two electrons alternate in one period between a rapid 
motion near the nucleus and a slow screened motion at a distance from 
the nucleus. 
Interestingly, the PO, which exhibits this alternation, is unstable.  
We found POs without binary collision through interpolation between 
two POs with binary collision. 
Some of stable frozen planetary orbits~\cite{RW} are  
among them (see Figs.~\ref{fig4}(e-g)).

Finally, we characterize POs as follows. 
In Figs.~\ref{fig4}(a-d), it is clear that one electron revolves 
round the nucleus more than the other. Its number of revolutions up to 
the turn-back point (in the half period) is characterized by a half 
integer or an integer, say $n_{r}$. 
For Figs.~\ref{fig4}(a-d), it is easily inferred that there exists a sequence 
of POs from $n_{r}=1$ to $n_{r}=\infty$ (i.e., toward ionization). 
For these POs, the Lyapunov exponents and the period are numerically calculated
(Fig.~\ref{fig5}).  
It shows a remarkable regularity of the period, 
$T=S/2=an_{r}+b$, where $T$ is the period and $S$ is the action.  

Now taking the partial summation over $n_{r}$ 
in the Gutzwiller periodic orbit sum 
$d_{\mbox{\scriptsize osc}}(E)=\sum_{p} A_{p} e^{\frac{i}{\hbar}S_{p}}$ 
with a daring approximation, i.e., $A_{p}$ is factored out and 
is replaced by the mean value $\overline{A}$, 
thanks to Poisson summation formula, 
we obtain the Rydberg energy levels, $E_{n}=-\frac{C}{n^{2}}$ 
with constant $C$. 
This means that the regular structure found is important 
for the energy levels of the helium atom. 
The amplitude factor $A_{p}$ may play a role of determining 
the quantum defects for the helium atom as observed in the collinear 
{\it eZe} case~\cite{TW}. 
To complete this program to obtain the energy levels, 
we have to seek the geometrical structure of the whole (unstable) POs. 
The result of the present Letter is a hint to this problem. 

In Fig.~\ref{fig6}, a schematic disposition of BCCs 
and POs are illustrated. 
As shown in Fig.~\ref{fig6}(a), 
there is a fundamental block between two BCCs of the second kind. 
The number of fundamental blocks is infinite (Fig.~\ref{fig6}(a)). 
The fractal structure of BCCs and the tree structure of POs 
similar to a binary Cayley tree are 
expected to be topologically the same in every block (Fig.~\ref{fig6}(b)) 
reflecting the fractal distribution of TCPs on $\alpha = \pi$. 
We show in Fig.~\ref{fig6}(b) the structure of the rightmost block. 
We find that there exists a PO on each lobe in Fig.~\ref{fig6}(b),  
even on the lobe of case (ii). 
It is interesting to note that though POs are far from 
$\alpha = \pi$, they keep a kind of one-dimensionality 
along with other orbits in BCCs. 

In conclusion, we have informally proved the existence of 
self-retracing POs with binary collisions, which form the backbone 
of the phase space, using the structure of BCCs and have indeed 
found such POs numerically.  
These findings would serve theoreticians with new stimuli, 
who would like to carry out the semiclassical treatment of 
the helium atom in two dimensions with zero angular momentum. 


This work was supported by a Grant-in-Aid for Scientific Research 
(No.17740252) from the MEXT, Japan.

\end{document}